\documentclass[twocolumn,aps,prd,superscriptaddress,nofootinbib]{revtex4-1}
\usepackage{amssymb}
\usepackage{mathbbol}
\usepackage{amsfonts}
\usepackage{mathrsfs}
\usepackage{epsfig,bm,dcolumn}
\usepackage{graphicx}
\usepackage{color}
\usepackage{amsmath}
\usepackage{dcolumn}
\usepackage{overpic}
\usepackage[colorlinks, linkcolor=red,anchorcolor=green,citecolor=blue]{hyperref}
\usepackage{slashed}

\usepackage[normalem]{ulem}

\usepackage{indentfirst}
\usepackage{bm}

\usepackage{simplewick}

\usepackage{xcolor}

\graphicspath{{picture/}}
\usepackage{color}

\begin{document}
	\title{The effect of vorticity on the dynamical magnetic fields in heavy-ion collisions}
	\author{Anping Huang}
	\affiliation{School of Material Science and Physics, China University of Mining and Technology, Xuzhou, China}
	\affiliation{School of Nuclear Science and Technology, University of Chinese Academy of Sciences, Beijing 100049, China}
	\affiliation{Physics Department, Tsinghua University, Beijing 100084, China}
 \email{}

\author{Xiang-Yu Wu}
\email{xiangyu.wu2@mail.mcgill.ca}

\affiliation{Department of Physics, McGill University, Montreal, QC, Canada H3A
2T8}
\affiliation{Department of Physics and Astronomy, Wayne State University, Detroit
MI 48201}
\affiliation{Institute of Particle Physics and Key Laboratory of Quark and Lepton Physics (MOE), Central China Normal University, Wuhan, 430079, China}

\author{Mei Huang} \email{huangmei@ucas.ac.cn}
\affiliation{School of Nuclear Science and Technology, University of Chinese Academy of Sciences, Beijing 100049, China}
\date{\today}

\begin{abstract}
Magnetic fields in heavy-ion collisions are pivotal and subject to diverse factors. In this study, we quantitatively investigate the impact of fluid vorticity on the evolution of magnetic fields in the 20-50\% centrality class in Au+Au collisions, with collision energies of $\sqrt{s_{NN}}=(7.7, 14.5, 19.6, 27, 39, 62.4, 200)$ GeV. Our results indicate that fluid vorticity leads to a delay in the evolution of the magnetic field, in which this effect becomes more pronounced as the collision energy decreases. Additionally, we have calculated the mean magnetic field values on the freeze-out hypersurface for various collision energies. Our simulation results align with the values inferred from experimental data of $\bar{\Lambda}-\Lambda$, within the error margins.
\end{abstract}
	
\maketitle

\section{Introduction}

In relativistic heavy-ion collisions, scientists create a minuscule yet intensely hot environment where quark-gluon plasma (QGP) forms. This extraordinary state of matter provides a unique window to probe the fundamental behavior of strongly interacting matter under extreme conditions, such as extremely high temperatures and ultra-high baryon densities. More recently, the intense magnetic fields produced in heavy-ion collisions have attracted significant interest due to their connection with various novel quantum effects. These include, but are not limited to, the Chiral Magnetic Effect, Chiral Magnetic Waves, directed flow characterized by $v_1$ of $D^0$ mesons, and the splitting of spin polarization observed in $\Lambda$ and $\bar{\Lambda}$ hyperons~\cite{Kharzeev:2012ph,Kharzeev:2015znc,Kharzeev:2020jxw,Fukushima:2018grm,Shovkovy:2021yyw,Li:2020dwr,Huang:2015oca}. Recently, the STAR Collaboration has for the first time observed the signal of the Electromagnetic Field Effect via the difference in charge-dependent directed flow in heavy-ion collisions~\cite{STAR:2023jdd}.

In non-central heavy-ion collisions, the high-speed motion of ions generates exceptionally powerful magnetic fields. In Au+Au collisions at the Relativistic Heavy Ion Collider (RHIC), the magnetic fields can achieve intensities on the order of $eB\sim\,m^{2}_{\pi}\sim\,10^{18}$~Gauss. At the Large Hadron Collider (LHC), these magnitudes can increase by an order of magnitude~\cite{Kharzeev:2007jp,Skokov:2009qp,Voronyuk:2011jd,Bzdak:2011yy,Deng:2012pc,Bloczynski:2012en,McLerran:2013hla,Tuchin:2015oka,Chen:2021nxs}.  Although the initial magnetic field is extremely strong, it decays rapidly. However, significant research indicates that the quark-gluon plasma can slow down the rate of decay due to its high electrical conductivity~\cite{Gupta:2003zh,Aarts:2007wj,Ding:2010ga,Ding:2016hua,Aarts:2002tn,Huang:2013iia,Jiang:2014ura,Bannur:2006js,Das:2019wjg,Hosoya:1983xm,Nam:2012sg,Cassing:2013iz,Arnold:2003zc,Wang:2021oqq}. In our earlier research \cite{Huang:2022qdn}, we conducted a comprehensive exploration of the dynamics of the magnetic field, employing the solution of Maxwell's equations against the hydrodynamics background in the context of $\sqrt{s_\mathrm{NN}}=200$~GeV Au+Au collisions. Our findings indicated that the dynamical magnetic field experiences suppression due to the longitudinal expansion of the QGP, leading to a weakened magnetic field in the final state.

It's noteworthy that the fluid vorticity generated by the non-zero angular momentum during non-central heavy-ion collisions can induce a new magnetic field, aligning with the vorticity's direction due to the swirling motion of charged particles within the QGP~\cite{Guo:2019mgh}. In this study, the authors initially estimated the magnitude of the induced magnetic field near the freeze-out surface using a phenomenological approach, primarily represented by the equation $eB=\frac{e^2}{4\pi}\,n_{q}A\,\omega$. Here, $n_{q}(\sqrt{S_{NN}})=0.3-0.087\ln(S_{NN})+0.0067(\ln(S_{NN}))^{2}$ defines the charge density at the freeze-out surface. Furthermore, $A=\pi R^{2}$ denotes the transverse area of the fluid vortex, with $R$ referring to the transverse size, and $\omega$ indicating the vorticity.  These induced magnetic fields might be stronger than previously assumed, potentially providing clarification for the observed discrepancies in spin polarization splitting between $\Lambda$ and $\bar{\Lambda}$ hyperons. 

In this study, we aim to delve deeper into the dynamic magnetic fields in heavy-ion collisions, paying particular attention to the effects of vorticity. The work will build upon our previous research~\cite{Huang:2022qdn}, considering the weak field method, where the influence of the medium on electromagnetic fields is taken into account while assuming the negligible back reaction of electromagnetic fields on the medium. In this approach, there is no need to solve the complete set of equations consistent with hydrodynamics and Maxwell's equations, similar to Magnetohydrodynamics (MHD)~\cite{Inghirami:2016iru, Inghirami:2019mkc}. 
The equations of hydrodynamics and Maxwell's equations can be separately solved. Consequently, the evolution of the medium can be described using standard viscous hydrodynamics without considering electromagnetic fields. Simultaneously, the evolution of the electromagnetic fields can be derived from Maxwell’s equations, taking into account responses from the medium, such as induction currents~\cite{Huang:2022qdn}.

The remainder of this paper is organized as follows. Section~\ref{sec-1} briefly introduces the framework involving Maxwell's equation and viscous hydrodynamics utilized in this study. Section~\ref{sec-2} explores the impact of vorticity on the dynamic evolution of magnetic fields in heavy-ion collisions with varying collision energies. Finally, Appendix~\ref{app-1} briefly outline the algorithm for solving Maxwell's equations in a hydrodynamic background.

\section{The numerical framework } \label{sec-1}
\subsection{The Maxwell equation in Milne space}

To quantitatively analyze the influence of fluid vorticity on the evolution of the magnetic field, we employ the weak field method, which builds upon prior research~\cite{Huang:2022qdn}. This method allows us to solve Maxwell's equations and viscous hydrodynamic equations independently. By utilizing this approach, we can investigate the relationship between fluid vorticity and magnetic field evolution more effectively. To facilitate the description of the evolution of the magnetic field within a fluid medium, Maxwell's equations are reformulated in Milne space as follows:

\begin{align}
	&\hat{D}_{\mu}F_{M}^{\mu\nu}=J^{\nu},\label{eq:maxm1}\\
	&\hat{D}_{\mu}\widetilde{F}_{M}^{\mu\nu}=0,\label{eq:maxm2}
\end{align} 
Here, the subscript $M$ is employed to denote the electromagnetic field tensor within Milne coordinates. The covariant derivative $\hat{D}_{\mu}$ acting on a tensor $t^{\nu\rho}$ is defined as:

\[
\hat{D}_{\mu}t^{\nu\rho}=\partial_{\mu}t^{\nu\rho}+\Gamma^{\nu}_{\lambda\mu}t^{\lambda\rho}+\Gamma^{\rho}_{\lambda\mu}t^{\nu\lambda},
\]

where $\Gamma^{\rho}_{\mu\nu}$ represents the Christoffel symbols, given by:

\[
\Gamma^{\rho}_{\mu\nu} = \frac{1}{2} g^{\rho\sigma}(\partial_{\nu}g_{\sigma\mu}+\partial_{\mu}g_{\sigma\nu}-\partial_{\sigma}g_{\mu\nu}).
\]

We adopt the metric convention $g_{\mu\nu}=\mathrm{diag}(1,-1,-1,-\tau^{2})$. Finally, the dual tensor is defined by:

\[
\widetilde{F}_{M}^{\mu\nu}=\frac{1}{2}\epsilon^{\mu\nu\rho\sigma}_{M}F^{M}_{\rho\sigma},
\]
where $\epsilon^{\mu\nu\rho\sigma}_{M}$ denotes the Levi-Civita symbol in Milne space. It is noteworthy that both the electromagnetic tensor and the Levi-Civita symbol differ from the Minkowski coordinate case. The explicit forms of these symbols can be found in Appendix~\ref{ap:levi-civita}.

The currents are composed of normal currents, diffusion current, Ohm's law, and CME current as the following,   
\begin{align}\begin{split}\label{eq:j}
		&J^{\mu}=J^{\mu}_{m}+J^{\mu}_{s},\\
		&J^{\mu}_{m}=n_{q}\,u^{\mu}+d_{q}^{\mu}+\sigma\,F^{\mu\nu}_{M}u_{\nu}+\sigma_{\chi}\widetilde{F}^{\mu\nu}_{M}u_{\nu}.	
\end{split}\end{align} 
Here, $J^{\mu}_{s}$ denotes the source contributions from fast-moving charged particles in heavy-ion collisions. $J^{\mu}_{m}$ represents the electric current in the medium, where $n_q$ is the charge number density, and $d_{q}^\mu$ represents the electric diffusive current. Additionally, $u^{\mu}$ denotes the fluid velocity of the QGP. The symbols $\sigma$ and $\sigma_{\chi}$ correspond to the electric conductivity and chiral conductivity of the QGP, respectively. In our simulation, we will set $\sigma_{\chi}$ to zero as its contribution is not considered significant. The electrical conductivity $\sigma$ is proportional to temperature, following the formula $\sigma/T=0.4$, which falls within the reasonable range estimated in prior work~\cite{Huang:2022qdn}. In the current study, we refrain from providing an in-depth exploration of electric conductivity due to its unresolved nature. Interested readers are directed to the comprehensive discussion on electric conductivity in the previously work~\cite{Huang:2022qdn}. 

For an intuitive understanding, we can conduct a simple qualitative analysis of the mechanism by which fluid vorticity induces a magnetic field within the framework of Maxwell's equations. To simplify our analysis, we will omit all terms in the current density equation Eq.(\ref{eq:j}) except for the first term. By transforming Eq.(\ref{eq:maxm1}) to Minkowski coordinates, we can derive the relationship between the induced magnetic field and fluid vorticity as follows:
\begin{align*}
&n_{q} \omega^{\mu}=\frac{1}{2}n_{q}\epsilon^{\mu\nu\alpha\beta}u_{\nu}\partial_{\alpha}u_{\beta}\approxeq(\partial_{\alpha}\partial^{\alpha}\widetilde{F}^{\mu\nu})u_{\nu}.
\end{align*}
In the local rest frame of a fluid cell and under the assumption of a static case, we have the relation $n_{q}\boldsymbol{\omega}=(\partial_{\alpha}\partial^{\alpha})\mathbf{B}$, which indicates that $\mathbf{B}$ is proportional to $n_{q}\,\boldsymbol{\omega} $. As mentioned in the introduction, the authors provide a qualitative relation between the induced magnetic field and fluid vorticity in their work\cite{Guo:2019mgh}, expressed as $eB=\frac{e^2}{4\pi}\,n_{q}A\,\omega$. In this context, the scenarios where $n_q=0$ and $n_q\neq 0$ are effectively interpreted as cases without and with vorticity contributions, respectively. Therefore, throughout this paper, we will refer to the scenarios without vorticity as corresponding to $n_q=0$, and those with vorticity as corresponding to $n_q\neq 0$ for intuitive clarity. Undoubtedly, the current analysis serves as a coarse qualitative approach. To delve deeper into the subject, it is essential to conduct simulations of the Maxwell equations to examine the influence of fluid vorticity on the magnetic field’s dynamics.

Next, Let's distinguish the electromagnetic fields into internal and external fields, following the approach introduced by McLerran and Skokov~\cite{McLerran:2013hla}, i.e
\[
F^{\mu\nu}_{M}=F^{\mu\nu}_{M, int}+F^{\mu\nu}_{M, ext},~~~~~~\widetilde{F}^{\mu\nu}_{M}=\widetilde{F}^{\mu\nu}_{M, int}+\widetilde{F}^{\mu\nu}_{M, ext}.
\]
The subscript "ext" signifies the external component originating from the source contribution of fast-moving charged particles in heavy-ion collisions, while "int" denotes the induced electromagnetic fields generated within the formed quark-gluon plasma (QGP). Subsequently, Eq. (\ref{eq:maxm1}) and Eq. (\ref{eq:maxm2}) can now be divided into two segments. Concerning the external component,
\begin{align}
	&\hat{D}_{\mu}F_{M, ext}^{\mu\nu}=J^{\nu}_{s},\\
	&\hat{D}_{\mu}\widetilde{F}_{M, ext}^{\mu\nu}=0.
\end{align}
Here, $J^{\nu}_{s}$ represents the source contributions originating from the fast-moving protons in the colliding nuclei. 
An analytical solution exists for this set of equations, providing the electric and magnetic fields induced by fast-moving charged particles. In this study, we consider the target nucleus as an optical sphere and apply Gauss's theorem in the rest frame of the target. Subsequently, we boost the electromagnetic fields to the laboratory frame using the velocity of the target. Further details can be found in previous work \cite{Huang:2022qdn}.

The internal part is written as 
\begin{align}\label{eq:maxin}
	\begin{split}
	&\hat{D}_{\mu}F_{M, int}^{\mu\nu}=J^{\nu}_{m},\\
	&\hat{D}_{\mu}\widetilde{F}_{M, int}^{\mu\nu}=0,\\
	&J^{\mu}_{m}=n_{q}\,u^{\mu}+d_{q}^{\mu}+\sigma\,(F^{\mu\nu}_{M,int}+F^{\mu\nu}_{M,ext})u_{\nu}\\
	&~~~~~~+\sigma_{\chi}(\widetilde{F}^{\mu\nu}_{M,int}+\widetilde{F}^{\mu\nu}_{M,ext})u_{\nu}.
\end{split}
\end{align}
With the following definitions of the electric field and magnetic field,
\begin{align}
	&\widetilde{E}^{i}=F_{M}^{i0},\qquad \widetilde{B}^{i}=\widetilde{F}_{M}^{i0}.
\end{align}
More explicitly, the electromagnetic tensors in Milne space are expressed in terms of the electric and magnetic fields as follows:
\begin{align*}
	&F^{\mu\nu}_{M}=\begin{pmatrix}
		0 & -\widetilde{E}^{x} & -\widetilde{E}^{y} & -\widetilde{E}^{\eta}\\
		\widetilde{E}^{x} & 0 & -\tau\,\widetilde{B}^{\eta} &\frac{\widetilde{B}^{y}}{\tau}\\
		\widetilde{E}^{y} &\tau\,\widetilde{B}^{\eta} & 0 & -\frac{\widetilde{B}^{x}}{\tau} \\
		\widetilde{E}^{\eta} & -\frac{\widetilde{B}^{y}}{\tau} &\frac{\widetilde{B}^{x}}{\tau} &0 \\
	\end{pmatrix}, \\	&\widetilde{F}^{\mu\nu}_{M}=\begin{pmatrix}
		0 & -\widetilde{B}^{x} & -\widetilde{B}^{y} & -\widetilde{B}^{\eta}\\
		\widetilde{B}^{x} & 0 & \tau\,\widetilde{E}^{\eta} &-\frac{\widetilde{E}^{y}}{\tau}\\
		\widetilde{B}^{y} &-\tau\,\widetilde{E}^{\eta} & 0 & \frac{\widetilde{E}^{x}}{\tau} \\
		\widetilde{B}^{\eta} & \frac{\widetilde{E}^{y}}{\tau} &-\frac{\widetilde{E}^{x}}{\tau} &0 \\
	\end{pmatrix}
\end{align*}

The set of equations in the internal part can be reformulated as follows,
\begin{align}
	\begin{split}
		&\partial_{x}\widetilde{E}^{x}+\partial_{y}\widetilde{E}^{y}+\partial_{\eta}\widetilde{E}^{\eta}=J^{\tau}_{m},\\
		&\partial_{\tau}(\tau\,\widetilde{E}^{x})=\partial_{y}(\tau^{2}\widetilde{B}^{\eta})-\partial_{\eta}\widetilde{B}^{y}-\tau\,J^{x}_{m},\\
		&\partial_{\tau}(\tau\,\widetilde{E}^{y})=-\partial_{x}(\tau^{2}\widetilde{B}^{\eta})+\partial_{\eta}\widetilde{B}^{x}-\tau\,J^{y}_{m},\\
		&\partial_{\tau}(\tau\,\widetilde{E}^{\eta})=\partial_{x}\widetilde{B}^{y}-\partial_{y}\widetilde{B}^{x}-\tau\,J^{\eta}_{m}.
	\end{split}\label{eq:ME-Milne-E}\\~\nonumber\\
	\begin{split}
		&\partial_{x}\widetilde{B}^{x}+\partial_{y}\widetilde{B}^{y}+\partial_{\eta}\widetilde{B}^{\eta}=0,\\
		&\partial_{\tau}(\tau\,\widetilde{B}^{x})=-\partial_{y}(\tau^{2}\widetilde{E}^{\eta})+\partial_{\eta}\widetilde{E}^{y},\\
		&\partial_{\tau}(\tau\,\widetilde{B}^{y})=\partial_{x}(\tau^{2}\widetilde{E}^{\eta})-\partial_{\eta}\widetilde{E}^{x},\\
		&\partial_{\tau}(\tau\,\widetilde{B}^{\eta})=-\partial_{x}\widetilde{E}^{y}+\partial_{y}\widetilde{E}^{x}.
	\end{split}\label{eq:ME-Milne-B}
\end{align} 
A carefully designed numerical algorithm is employed to solve these above two set of equations. The Yee-grid algorithm is chosen for its robust stability and efficiency, as demonstrated in prior studies~\cite{McLerran:2013hla, Zakharov:2014dia, Huang:2022qdn}. A brief summary of this algorithm is presented in Appendix~\ref{app-1}. When presenting the numerical results, we will transform the electric and magnetic fields to the Minkowski coordinate case using the following expressions,
\begin{align}\begin{split}
		E^{x}=\;&\cosh\eta\,\widetilde{E}^{x}+\sinh\eta\,\widetilde{B}^{y}, \\
		E^{y}=\;&\cosh\eta\,\widetilde{E}^{y}-\sinh\eta\,\widetilde{B}^{x},\\
		E^{z}=\;&\tau\widetilde{E}^{\eta},\\
		B^{x}=\;&\cosh\eta\,\widetilde{B}^{x}-\sinh\eta\,\widetilde{E}^{y},\\
		B^{y}=\;&\cosh\eta\,\widetilde{B}^{y}+\sinh\eta\,\widetilde{E}^{x},\\
		B^{z}=\;&\tau\widetilde{B}^{\eta}.
\end{split}\end{align}

To investigate the impact of vorticity on the magnetic field, it is essential to consider the first and second terms of the medium current in Eq. (\ref{eq:j}). Notably, these terms were previously set to zero in the study~\cite{Huang:2022qdn}. In this study, the electric charge density is expressed as $n_Q=0.4 e\,n_B$, where the factor $0.4$ is determined by the ratio of the number of nucleons to protons in Au-Au collisions. Here, $e$ represents the element charge, and $n_B$ denotes the number density of baryons. Regarding the electric diffusion current $d^{\mu}_q$, it can be set equal to 0.4 times the elementary charge $e$ multiplied by the baryon diffusion current, that is, $d^{\mu}_Q=0.4 e\,d^{\mu}_B$. In reality, the impact of the electric diffusion current on the magnetic field is considered negligible; therefore, we will not address it in subsequent discussions.. It is now clearer what background information is required by the above Maxwell's equations to evolve the electric and magnetic fields in Eq. (\ref{eq:maxm1}) and Eq. (\ref{eq:maxm2}). The essential quantities include the fluid velocity $u^{\mu}$, the temperature  $T$ of the QGP, and either the baryon number density or the baryon current. In this study, we employ CLVisc hydrodynamics to calculate these quantities, which consistently incorporates both vorticity and baryon information.

\subsection{The hydrodynamics evolution of QGP background at finite baryon density}
The dynamic evolution of the QGP medium is described by the (3+1)-dimensional CLVisc hydrodynamics framework\cite{Pang:2012he,Pang:2018zzo,Wu:2021fjf}. This approach has been successfully applied to theoretically calculate the bulk properties of QGP medium, as well as polarization and helicity of the $\Lambda$ hyperon across a wide range of collision energies at RHIC experiments\cite{Wu:2021fjf,Jiang:2023fad,Wu:2022mkr,Yi:2023tgg}. 

In this work, the state-of-art SMASH hadronic transport model\cite{Schafer:2021csj,SMASH:2016zqf,Gotz:2023kkm,Gotz:2022naz,Inghirami:2022afu} was chosen to simulate the initial conditions and the subsequent out-of-equilibrium evolution in heavy ion collisions.  SMASH is initially designed to simulate the non-equilibrium microscopic motions of hadrons at lower collision energies. By considering the processes of elastic collisions, resonance formation and decay, as well as string fragmentation for all hadrons with masses up to 2.35 GeV, SMASH effectively provides a solution to the relativistic Boltzmann equation through the Monte Carlo method. More recently, the SMASH community has extended the SMASH code to include the initial condition module for heavy ion collisions. The initial protons and neutrons are sampled from the nuclear Woods-Saxon distribution and propelled with beam momentum until the first initial collision occurs. Subsequently, these initial particles undergo the aforementioned collision processes until they or newly produced hadrons approach the hypersurface, at which point the proper time equals the initial hydrodynamic starting time, denoted as $\tau_0$. At the hypersurface with the initial proper time $\tau_0$, the initial energy-momentum tensor and baryon current can be constructed based on the assumption that all hadrons have reached local thermal equilibrium as follows.
\begin{equation}
\begin{aligned}
T^{\mu \nu}\left(\tau_0, x, y, \eta\right) & = \sum_i \frac{p_i^\mu p_i^\nu}{p_i^\tau} G\left(\tau_0, x, y, \eta\right), \\
J^\mu\left(\tau_0, x, y, \eta\right) & =\sum_i B_i \frac{p_i^\mu}{p_i^\tau} G\left(\tau_0, x, y, \eta\right),
\end{aligned}
\end{equation}
Here $G\left(\tau_0, x, y, \eta\right)$ is the Gaussian smearing kernel function relative to the grid point $(x,y,\eta)$
\begin{equation}
\begin{aligned}
& G\left(\tau_0, x, y, \eta\right) \\
& \quad=\frac{1}{\mathcal{N}} \exp \left[-\frac{\left(x-x_i\right)^2+\left(y-y_i\right)^2}{2 \sigma_r^2}-\frac{\left(\eta-\eta_{s i}\right)^2}{2 \sigma_{\eta}^2}\right]
\end{aligned}
\end{equation}
and $p_i^\mu$
\begin{equation}
p^\mu=\left[m_T \cosh \left(Y-\eta_s\right), p_x, p_y, \frac{1}{\tau_0} m_T \sinh \left(Y-\eta_s\right)\right]
\end{equation}
is the four-momentum of hadrons at Melin coordinate with transverse mass $m_T$, rapidity $Y$, and space-time rapidity $\eta_s$. $B_i$ is the baryon number relative to hadrons and $\mathcal{N}$ is the normalization factor to keep the energy-momentum and baryon number conservation from SMASH transport approach to hydrodynamics calculation. In addition, the initial proper time $\tau_0$ is determined by the time it takes for the two colliding nuclei to completely pass through each other. It can be roughly estimated using the following formula,
\begin{equation}
\tau_0=\frac{2 R}{\sinh \left(y_{\text {beam}}\right)}
\end{equation}
which relates to different collision energies, the radius of the nucleus R and the beam rapidity beam
 $y_{\text {beam}}$. Finally, the Gaussian smearing width $\sigma_r$, $\sigma_{r}$ can be determined via comparing the hydrodynamics calculation to the experimental data on identified particle yield at the most central collisions. 

At the RHIC-BES collision energies, the effects of net-baryon density play a crucial role in the evolution of the QGP medium due to the stronger stopping power at lower collision energies. Therefore, the CLVisc hydrodynamics framework \cite{Pang:2012he,Pang:2018zzo,Wu:2021fjf} numerically solves both the energy-momentum and baryon number conservation equations according to the KT algorithm.
\begin{equation}
\begin{aligned}
\nabla_\mu T^{\mu \nu} & =0, \\
\nabla_\mu J^\mu_B & =0,
\end{aligned}
\end{equation}
The energy-momentum tensor and net-baryon current can be  decomposed as
\begin{equation}
\begin{aligned}
T^{\mu \nu} & =(e+P) u^\mu u^\nu-P g^{\mu \nu}+\pi^{\mu \nu}, \\
J^\mu_B & =n_B u^\mu+d^\mu_B,
\end{aligned}
\end{equation}
with energy density $e$, pressure $P$, flow velocity $u^{\mu}$, the shear-stress tensor $\pi^{\mu v}$, and baryon diffusion current $d^\mu_B$. In this work, we neglect the contribution of bulk pressure $\Pi$ to simplify the model calculations. The diffusion current  $\pi^{\mu\nu}$ and $u^\mu$ satisfy the Israel-Stewart second-order hydrodynamics equations. During hydrodynamic evolution, the specific shear viscosity is assumed to be independent of temperature but dependent on collision energy\cite{Karpenko:2015xea}. The baryon diffusion coefficient is set to 0. The NEOS-BQS equation of state (EOS)\cite{Monnai:2021kgu,Monnai:2019hkn} is also apply to close the system of hydrodynamics equations. The NEOS-BQS EOS is applied to close the system of hydrodynamic equations. To incorporate the chemical potential dependence into the EOS, it utilizes the HOTQCD lattice QCD EOS\cite{HotQCD:2014kol} at zero chemical potential and performs a Taylor expansion about finite chemical potential, based on lattice susceptibility calculations.  The QGP medium evolves until the local energy density $e_{\rm frz}$ drops to 0.4 $\rm GeV/fm^3$. Then thermodynamic variables and freezout volume element on the QGP freeze-out hypersurface can be extracted via Projection method\cite{Pang:2012he}.

In this paper, the smooth SMASH initial condition is generated by averaging 5000 event-by-event fluctuating SMASH initial events for a given centrality bin. Finally, the collision energy dependencies of free parameters in the hydrodynamic calculations are presented in a Table~\ref{tab:hydro}

\begin{table}
\begin{tabular}{|l|l|l|l|l|}
\hline$\sqrt{s_{\mathrm{NN}}}[\mathrm{GeV}]$ & $\tau_0[\mathrm{fm} ]$ & $\sigma_{r}[\mathrm{fm}]$ & $\sigma_{}[\mathrm{fm}]$ & $ C_{\eta}$ \\
\hline 7.7 & 3.2 & 1.0 & 0.35 & 0.2 \\
\hline 14.5  & 1.7 & 1.0 & 0.35 & 0.2 \\
\hline 19.6 & 1.2 & 1.0 & 0.35 & 0.15 \\
\hline 27  & 1.0 & 1.0 & 0.35 & 0.12 \\
\hline 39 & 0.9 & 1.0 & 0.5 & 0.08 \\
\hline 62.4 & 0.7 & 1.0 & 0.55 & 0.08 \\
\hline 200 & 0.4 & 1.0 & 0.8 & 0.08 \\
\hline
\end{tabular}
\caption{Collision energy dependence of the free parameters in hydrodynamics simulation.}
\label{tab:hydro}
\end{table}

\section{Numerical results}\label{sec-2}

Before presenting the numerical results, let us first introduce the relevant physical picture. As described in previous work~\cite{Huang:2022qdn}, the evolution of the QGP in heavy-ion collisions typically unfolds in three stages: the initial stage, the pre-equilibrium stage, and the hydrodynamic stage. Similarly, the evolution of the magnetic field can be divided into corresponding three stages. However, in this work, since our focus is on the contribution of the fluid vorticity field to the magnetic field, the evolution of the magnetic field is divided into only two stages: the pre-equilibrium stage and the hydrodynamic stage. In the the pre-equilibrium stage, the evolution of the electromagnetic field is assumed to occur in a vacuum, unaffected by the QGP medium. During the hydrodynamic stage, the Maxwell's equations are solved against the background of QGP medium. As a result, the evolution of the magnetic field is influenced by the properties of the QGP medium, especially by the fluid's vorticity field. This background information includes fluid velocity, temperature, baryon number density, and baryon number current density, among others.

To enable a comparison between the magnetic field derived from  Maxwell's equations and the magnetic field assessed from the $\bar{\Lambda}$-$\Lambda$ experimental measurement~\cite{Guo:2019mgh, Becattini:2020ngo, STAR:2017ckg}, the centrality bins are chosen to be  20-50\%. Based on the initial states module of the CLVisc model for Au+Au collision systems at energies of $\sqrt{s_{NN}} = (7.7, 14.5, 19.6, 27, 39, 62.4, 200)$ GeV, the corresponding impact parameters are $b = (8.873, 8.844, 8.856, 8.861, 8.846, 8.892, 8.889)$ fm, which are used to estimate the initial magnetic field. The initial proper times for the hydrodynamic simulations corresponding to these collision systems are listed in the second column of Table~\ref{tab:hydro}.

To ensure clarity in presenting the results, the following figures show the evolution of the magnetic field at the center of the fireball ($\mathbf{x} = (0, 0, 0)$). However, our framework can also provide the electric and magnetic fields at any arbitrary coordinates.

\begin{figure*}[!hbt]\centering
	\includegraphics[width=0.3\textwidth]{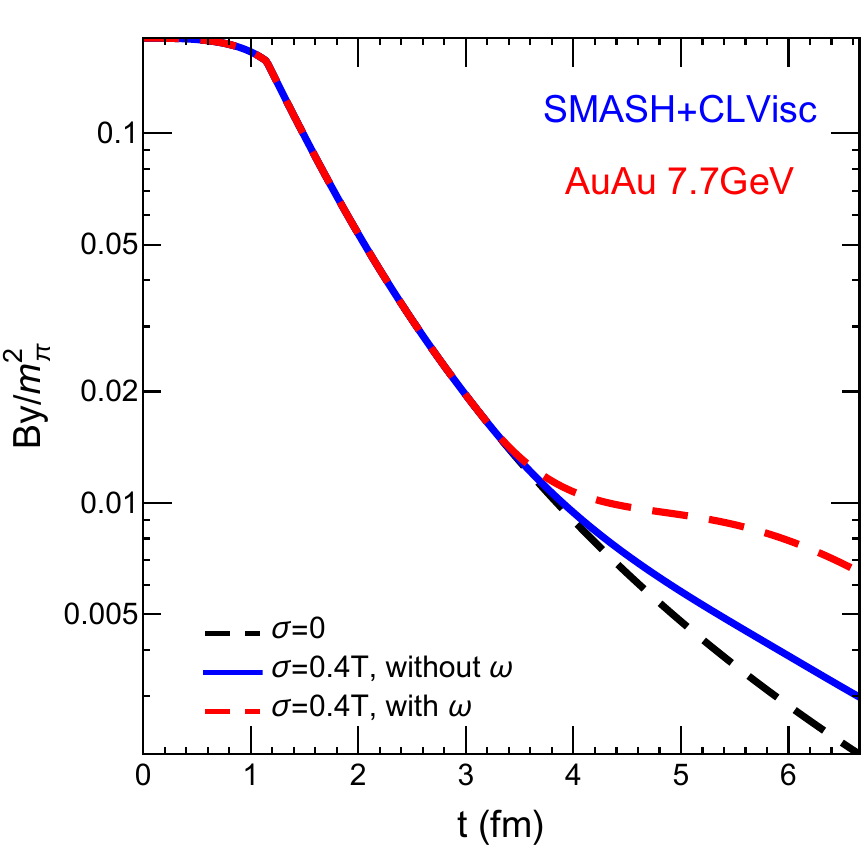}
	\includegraphics[width=0.3\textwidth]{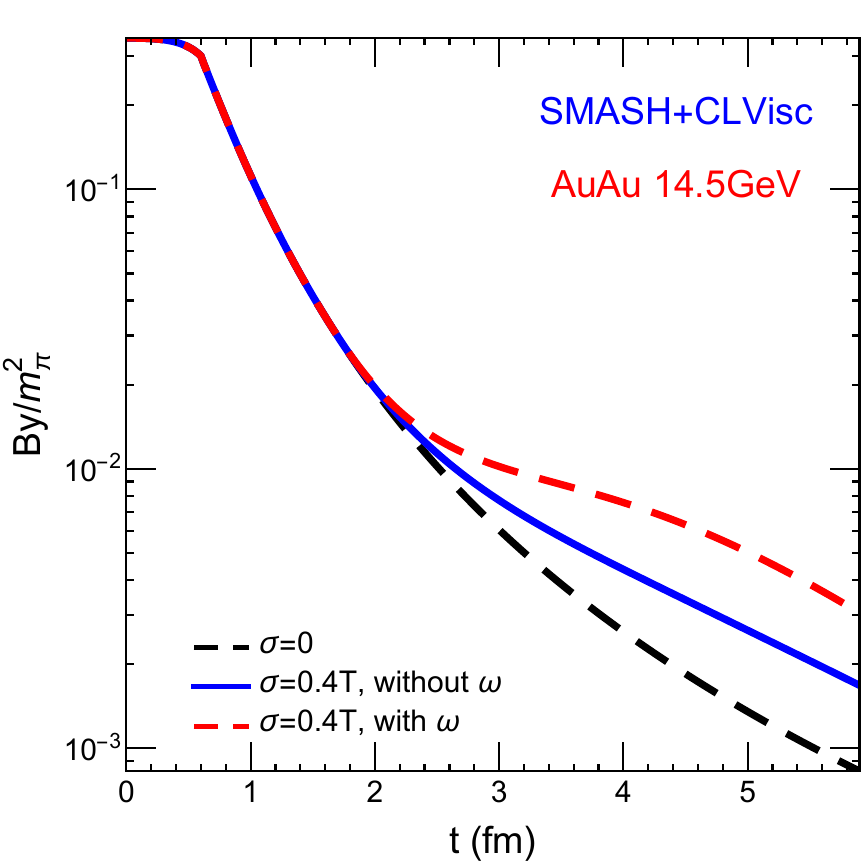}
	\includegraphics[width=0.3\textwidth]{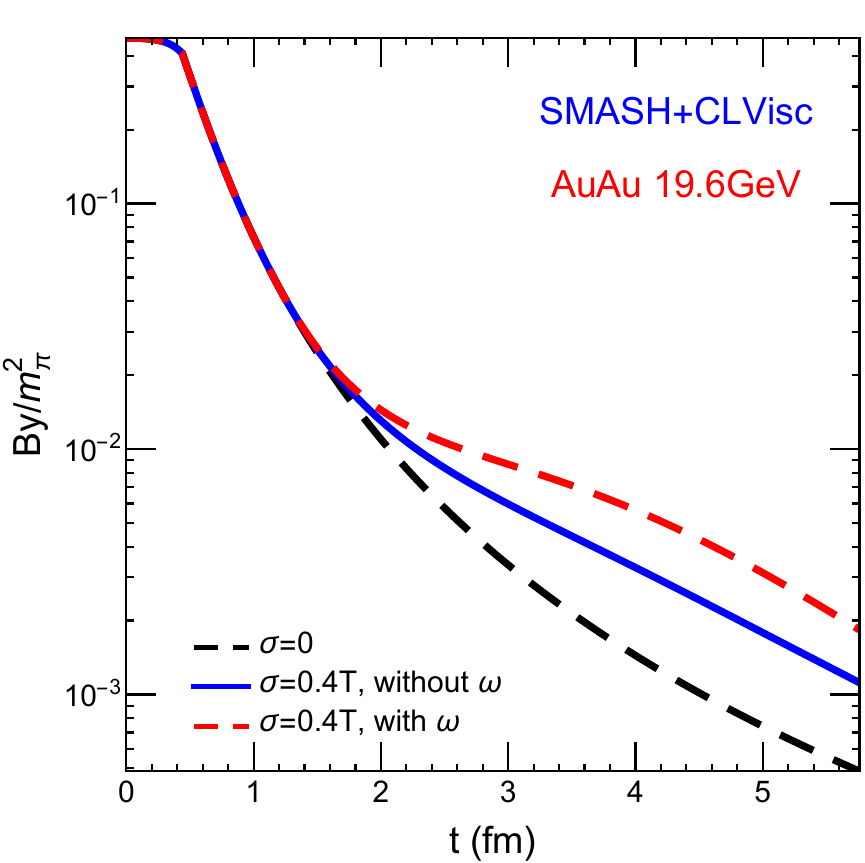}
	\includegraphics[width=0.3\textwidth]{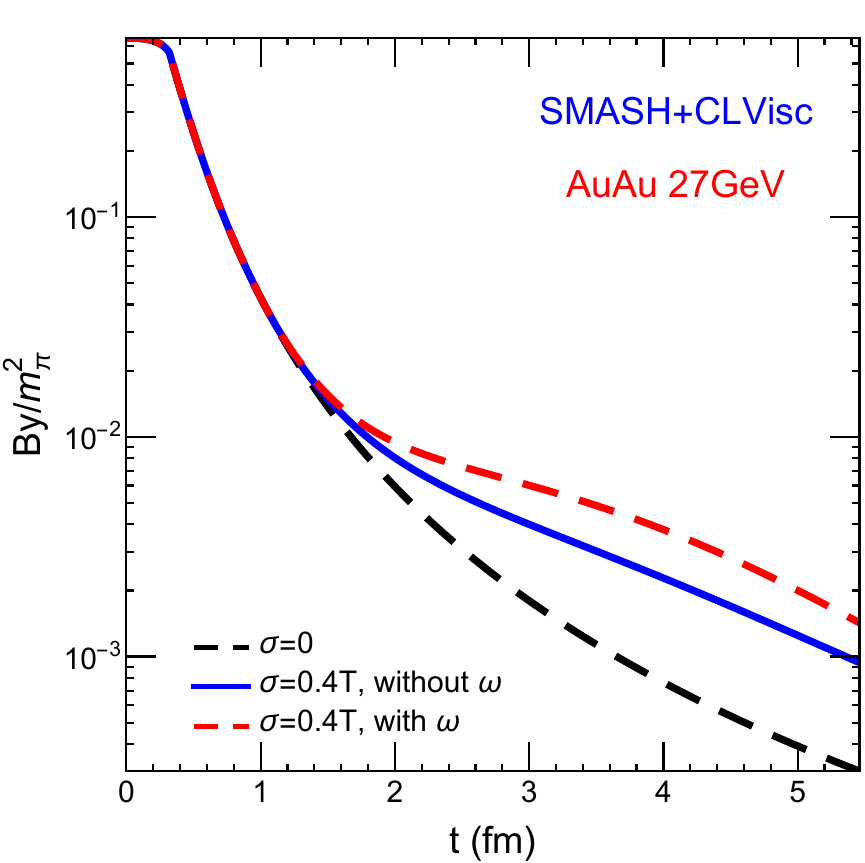}
	\includegraphics[width=0.3\textwidth]{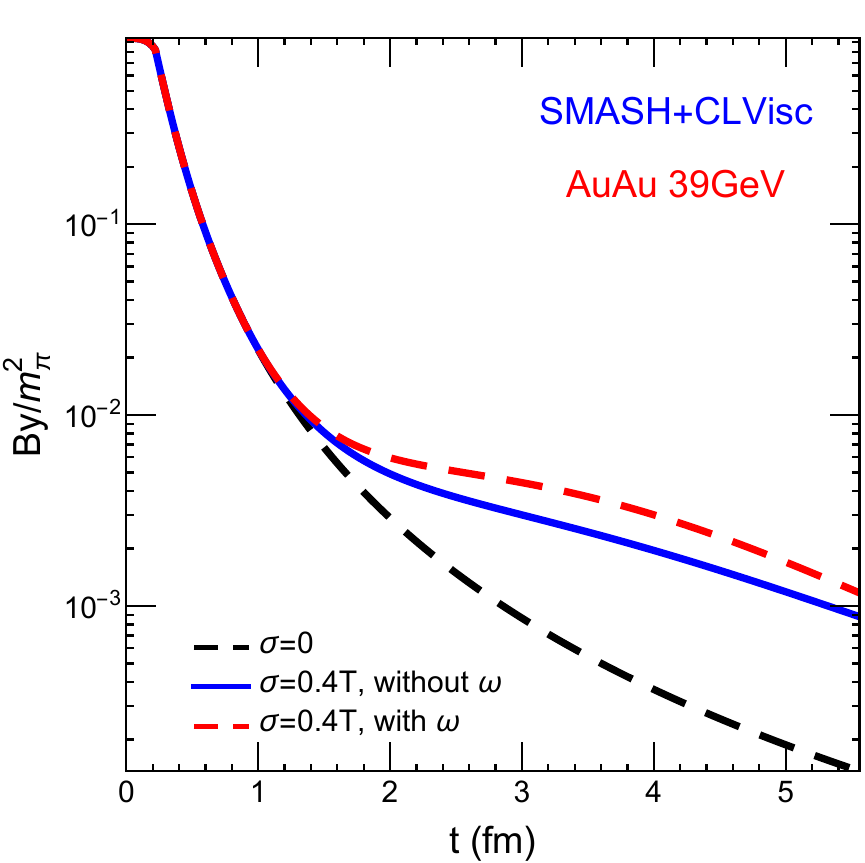}
	\includegraphics[width=0.3\textwidth]{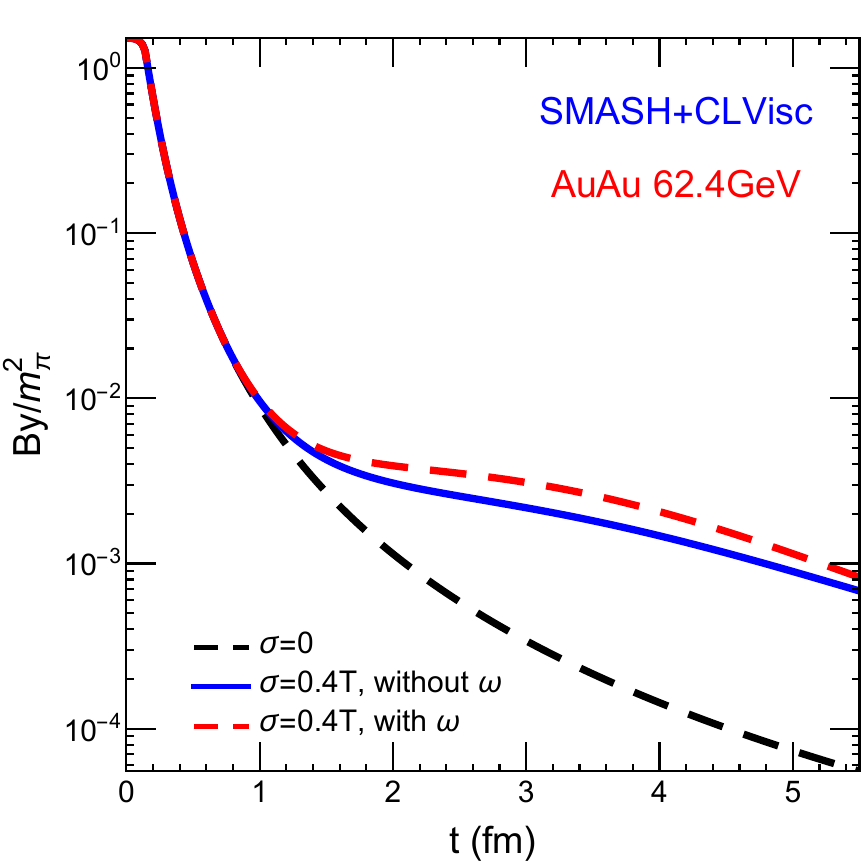}
	\includegraphics[width=0.3\textwidth]{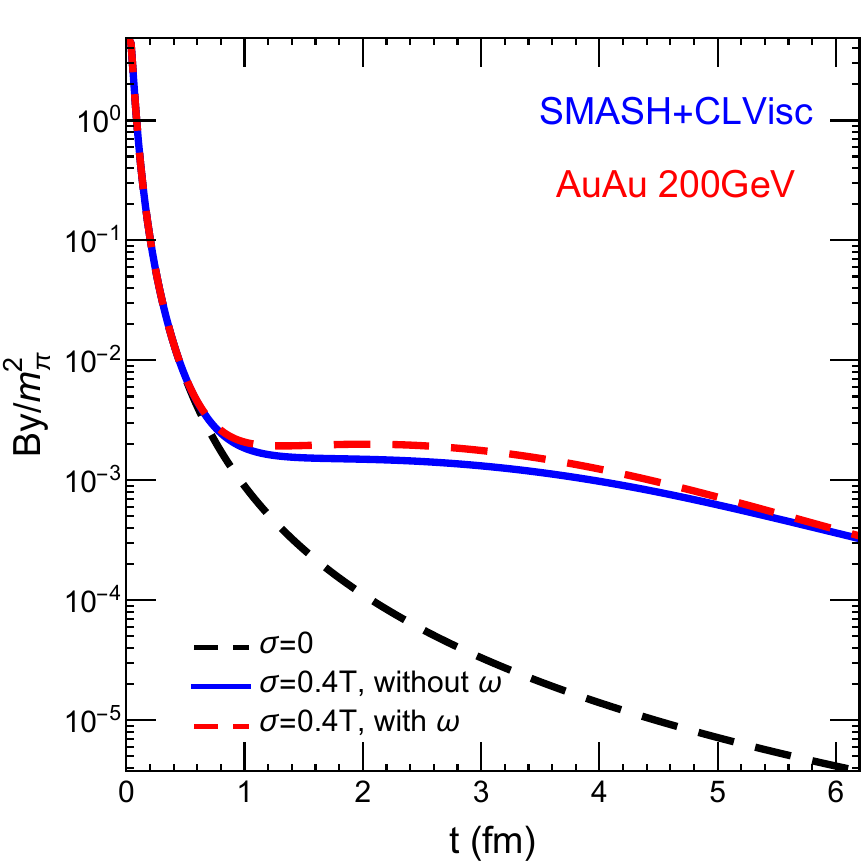}
	\caption{At the center of the fireball $(0, 0, 0)$, the evolution of the magnetic field over time is depicted, with the black dashed line representing the magnetic field's evolution in vacuum. The blue solid line and the red dashed line indicate the magnetic field within the medium, where the blue solid line is associated with a scenario where there is no vorticity, i.e., $\omega=0$, whereas the red dashed line corresponds to a situation where vorticity is present, i.e., $\omega\neq 0$. The collision energies are respectively $\sqrt{s_{NN}}=(7.7, 14.5, 19.6, 27, 39, 62.4, 200)$ GeV, with a centrality set at 20-50\%.\label{fig:omega-B}}
\end{figure*}

In Figure~\ref{fig:omega-B}, we have plotted the time evolution of the magnetic field in the 20-50\% centrality class Au+Au collision across collision energies of $\sqrt{s_{NN}}$=(7.7, 14.5, 19.6, 27, 39, 62.4, 200) GeV. The black dashed lines represent the magnetic field evolution in a vacuum. The blue solid lines and red dashed lines depict the field in a medium, without and with the vorticity effect, respectively.  It is evident that the medium somewhat slows the decay of the magnetic field within the fluid due to the contribution of electrical conductivity, which is consistent with previous studies \cite{Huang:2022qdn, Skokov:2009qp}. Next, we compare the results with and without the vorticity effect, it is found the vorticity within the QGP medium further significantly slows the decay of the magnetic field. Moreover, the effect of vorticity is initially enhanced at the early stages of hydrodynamic evolution and then gradually diminishes over time. Let us focus on the collision energy, it is noteworthy that the fluid's vorticity plays a key role in the evolution of magnetic field at lower collision energy. This effect is especially pronounced at $\sqrt{s_{NN}}$ = 7.7 GeV Au+Au collision energy, where the influence of vorticity substantially exceeds that of electrical conductivity. As the collision energy increases, the impact of the fluid's vorticity on the magnetic field gradually decreases, becoming negligible at $\sqrt{s_{NN}}$ = 200 GeV collision energy.  Notably, the fluid's vorticity decreases proportionally as the collision energy increases.  This trend is consistent with the results of previous works\cite {STAR:2017ckg, Jiang:2016woz}. Moreover, the strength of vorticity fluid weakens over time as well. It implies a direct proportional relationship between the magnetic field $B$ and the vorticity $\omega$ of the fluid,  $B\propto\omega$. This relationship is consistent with the formula derived from the work of Guo et al., $eB=\frac{e^2}{4\pi}\,n_{q}A\,\omega$\cite{Guo:2019mgh}.

\begin{figure*}[!hbt]
\centering
	\includegraphics[width=0.5\textwidth]{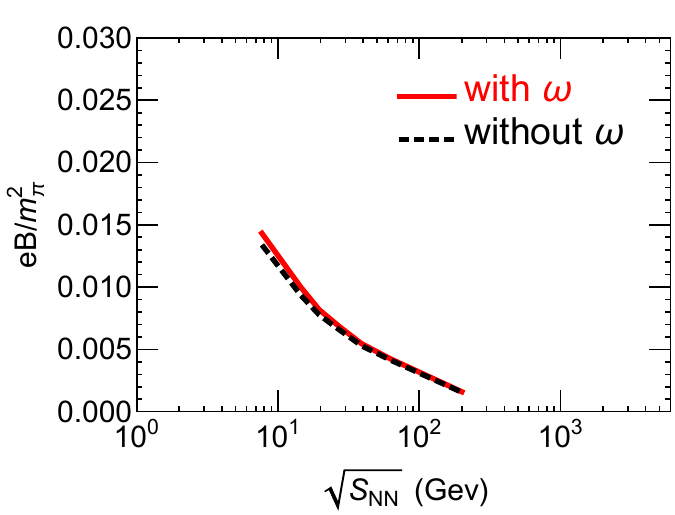}
	\caption{The strength of the magnetic field on the freeze-out hypersurface. The collision energies are set at $\sqrt{s_{NN}}=(7.7, 14.5, 19.6, 27, 39, 62.4, 200)$ GeV, with a centrality of $20-50\%$ and space-time rapidity $|\eta_s|\leq 1$.\label{fig:w-o-omega}}
\end{figure*}

To further evaluate the impact of vorticity on the magnetic field, we calculated the magnetic field's magnitude along y direction at the freeze-out hypersurface of the QGP fireball using the formula $\int d\sigma^{\mu}u_{\mu}B_{y}/\int d\sigma^{\mu}u_{\mu}$, where $d\sigma^{\mu}$ represents the element of hypersurface area. 
This estimation was considered both with and without the effects of vorticity. Figure~\ref{fig:w-o-omega} presents our estimation as a function of collision energy in the 20-50\% centrality bins and space-time rapidity windows $|\eta_s| \le 1$. The red solid line illustrates the magnetic field contribution incorporating vorticity, whereas the black dashed line denotes the contribution from electrical conductivity alone, excluding the effects of the vorticity field. It is found that the contribution of vorticity to the magnetic field on the hypersurface appears to be relatively small, with no evident enhancement. 
This is due to the relatively late initiation time of our fluid, resulting from the effect of finite overlap size at RHIC-BES energies, as shown in Figure \ref{fig:omega-B}. By this time, the magnetic field has decayed to a small value, thus limiting the contribution of the vorticity field. 

\begin{figure*}[!hbt]
\centering
	\includegraphics[width=0.5\textwidth]{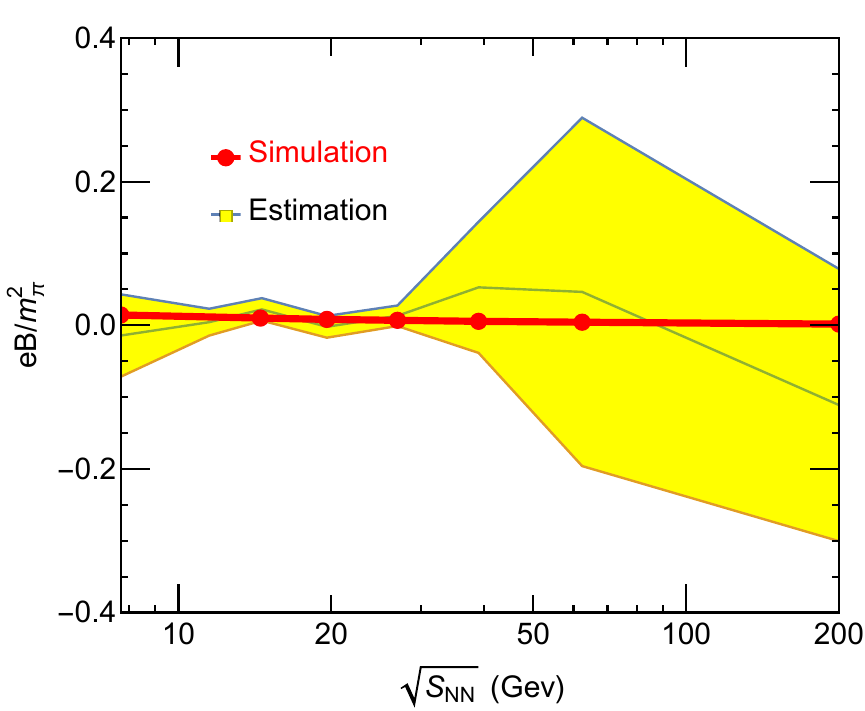}
	\caption{Comparison of the magnetic field strengths on the freeze-out hypersurface, where the red solid line represents the results from simulation calculations, and the yellow region corresponds to the magnetic field strengths estimated from the experimental data of $\bar{\Lambda}$-$\Lambda$. The respective collision energies are $\sqrt{s_{NN}}=(7.7, 14.5, 19.6, 27, 39, 62.4, 200)$ GeV, centrality is set at $20-50\%$, and rapidity $|y|\leq 1$.\label{fig:compare}}
\end{figure*}

On the other hand, assuming that the spin polarization splitting of $\bar{\Lambda}$/$\Lambda$ hyperons is caused by the magnetic field, we can assess the magnitude of the magnetic field on the freeze-out hypersurface of QGP fireball in heavy-ion collisions using experimental data and the formula $eB\approx\frac{M_N T}{0.613}(P_{\bar{\Lambda}}-P_{\Lambda})$ , where $M_N=0.938$ GeV \cite{Guo:2019mgh,Xu:2022hql}. To maintain consistency with previous studies, the temperature $T$ in this function will be adjusted using a fitted form given by $T=0.158-0.14\mu_B^{2}-0.04\mu_B^{4}$, where $\mu_B$ represents the baryon chemical potential. The baryon chemical potential is expressed as a function of the collision energy as $\mu_B=1.477/(1+0.343\sqrt{s_{NN}})$, with all quantities reported in GeV units\cite{Luo:2017faz, Li:2018ygx}. Here, we assume the magnetic field is along the y-axis to facilitate direct comparison with the simulated magnetic fields. In Figure \ref{fig:compare}, we compare the simulation results with the assessed values, specifically focusing on the magnitude of the magnetic field on the freeze-out hypersurface. The red solid line represents the results of the simulation calculations, while the yellow range indicates the estimated magnetic field strength based on the experimental data of $\bar{\Lambda}$-$\Lambda$, considering only the statistical errors of the experimental data. 
Here, we would like to note that the data for collision systems at $\sqrt{s_{NN}} = (39, 62.4, 200)$ GeV are sourced from \cite{STAR:2017ckg}, while the data for $\sqrt{s_{NN}} = (19.6, 27)$ GeV are obtained from \cite{STAR:2023nvo}. The remaining data are from STAR preliminary results\cite{STAR:huqiang}. All of these data for centrality are within the range of 20-50\% and mid-rapidity  $|y|\leq 1$.  
It should be noted that the type of rapidity $y$ mentioned here is different from the rapidity used to estimate the magnetic field in the Maxwell simulation. However, we assume that the $\Lambda$ and $\bar{\Lambda}$ hyperons mainly originate from the central space-time rapidity of the QGP medium and are influenced by the magnetic field in the same region.
As depicted in Figure \ref{fig:compare}, the magnetic fields obtained through simulations are generally in agreement with the ranges derived from experimental data. Moreover, the magnetic field's magnitude is small, but it is not zero. This indicates that accurately detecting the magnetic field signal through the spin polarization splitting of $\bar{\Lambda}$/$\Lambda$ hyperons is a significant challenge. In the future, it will be important to explore QGP signals that are more sensitive to the magnetic field, such as directed flow.


\section{Conclusion} \label{sec-3}

We have conducted a quantitative study on the contribution of fluid vorticity to the evolution of the magnetic field, employing the weak-field approximation method to solve the fluid dynamics and Maxwell's equations separately. The fluid dynamics component utilized the CLVisc hydrodynamics framework and the initial vorticity field is provided by the SMASH initial condition. The fluid dynamics component offers background information for each spacetime point, such as the fluid's velocity, temperature, baryon number density, and baryon number flux density etc. Maxwell's equations are then solved against this background.

The evolution of the magnetic field in the fireball resulting from Au+Au collisions in  20-50\% centrality class, across various collision energies: $\sqrt{s_{NN}}=(7.7, 14.5, 19.6, 27, 39, 62.4, 200)$ GeV, has been studied. 

It has been observed that the effect of fluid vorticity on the evolution of the magnetic field grows more significant as the collision energy decreases. This effect is particularly evident at lower energies, driven by an increase in the fluid's vorticity as the collision energy is reduced. When collision energy is fixed, the contribution of the vorticity on the magnetic field is most pronounced at the early stage of the fluid phase. However, this contribution diminishes over time as the initially high vorticity gradually decreases.
Comparisons were also made between the magnetic field on the freeze-out hypersurface as obtained from simulations and the values deduced from experimental data of $\bar{\Lambda}-\Lambda$. The simulation results generally fall within the error range of the deduced values. However, several factors still need to be addressed in this calculation. Firstly, the fluid dynamics in our simulations are initiated at a notably late stage, specifically at 3.2 fm/c for the 7.7 GeV collision system, as listed in the second column of Table~\ref{tab:hydro}. By this time, the magnetic field has significantly decayed to a small value, resulting in a corresponding decrease in the magnetic field strength on the freeze-out hypersurface. As previously stated, the current study does not incorporate the medium's influence on the magnetic field during the pre-equilibrium phase. Secondly, in our calculations, the charge density is set to $n_Q=0.4e\,n_B$, and the potential redistribution of charge density under the influence of electromagnetic fields was not considered. This could contribute to the observed discrepancy. 


In the future, to address the factors mentioned above and to better represent the evolution of the magnetic field in heavy-ion collisions, we plan to develop new approaches that incorporate contributions from the pre-equilibrium stage. During this stage, the QGP fireball is in a non-equilibrium state, making the conductivity and vorticity of the medium difficult to define. Therefore, one potential approach is to utilize transport theories, such as the Vlasov transport equation, to study the medium's effect on the evolution of the magnetic field. Additionally, we will incorporate the conservation of electric current density, $\partial_{\mu}J^{\mu}_q=0$, into the our Maxwell equations code to account for the redistribution of charge density under the influence of electromagnetic fields.

\section*{Acknowledgments}
The research of AH is supported by the National Natural Science Foundation of China (NSFC) Grant Nos.12205309. AH and MH are supported by the Strategic Priority Research Program of Chinese Academy
of Sciences Grant No. XDB34030000, the Fundamental Research Funds for the Central Universities, and the National Natural Science Foundation of China (NSFC) Grant Nos. 12235016.
MH is also supported in part by the National Natural Science Foundation of China (NSFC) Grant Nos.11725523, 11735007, and 12221005. 
X.-Y.W. is supported in part by the National Science Foundation (NSF) within the framework of the JETSCAPE collaboration under grant number OAC-2004571 and in part by the Natural
Sciences and Engineering Research Council of Canada (NSERC).

\begin{appendix}
\begin{widetext}
\section{The numerical algorithm of Eq.(\ref{eq:ME-Milne-E}) and Eq.(\ref{eq:ME-Milne-B})}\label{app-1}
The algorithms corresponding to Eq. (\ref{eq:ME-Milne-E}) can be expressed as follows:
\begin{align}
	\begin{split}
		&\partial_{\tau}(\tau\,\widetilde{E}^{x})|^{n}=\partial_{y}(\tau^{2}\widetilde{B}^{\eta})|^{n}-\partial_{\eta}\widetilde{B}^{y}|^{n}-\tau\,J^{x}_{m}|^{n},\\
		&\partial_{\tau}(\tau\,\widetilde{E}^{y})|^{n}=-\partial_{x}(\tau^{2}\widetilde{B}^{\eta})|^{n}+\partial_{\eta}\widetilde{B}^{x}|^{n}-\tau\,J^{y}_{m}|^{n},\\
		&\partial_{\tau}(\tau\,\widetilde{E}^{\eta})|^{n}=\partial_{x}\widetilde{B}^{y}|^{n}-\partial_{y}\widetilde{B}^{x}|^{n}-\tau\,J^{\eta}_{m}|^{n}.
	\end{split}
\end{align}
The symbol $|^{n}$ signifies that the corresponding function is defined at time $\tau = \tau_0 + n d\tau$. Employing the center difference method for time derivatives, the aforementioned set of equations can now be reformulated as follows,
\begin{align}
	\begin{split}
		&\widetilde{E}^{x}|^{n+1/2}=\frac{\tau_{n-1/2}}{\tau_{n+1/2}}\widetilde{E}^{x}|^{n-1/2}+\frac{d\tau}{\tau_{n+1/2}}\left[\partial_{y}(\tau^{2}\widetilde{B}^{\eta})-\partial_{\eta}\widetilde{B}^{y}\right]|^{n}-d\tau\frac{\tau_{n}}{\tau_{n+1/2}}\,J^{x}_{m}|^{n},\\
		&\widetilde{E}^{y}|^{n+1/2}=\frac{\tau_{n-1/2}}{\tau_{n+1/2}}\widetilde{E}^{y}|^{n-1/2}-\frac{d\tau}{\tau_{n+1/2}}\left[\partial_{x}(\tau^{2}\widetilde{B}^{\eta})-\partial_{\eta}\widetilde{B}^{x}\right]|^{n}-d\tau\frac{\tau_{n}}{\tau_{n+1/2}}\,J^{y}_{m}|^{n},\\
		&\widetilde{E}^{\eta}|^{n+1/2}=\frac{\tau_{n-1/2}}{\tau_{n+1/2}}\widetilde{E}^{\eta}|^{n-1/2}+\frac{d\tau}{\tau_{n+1/2}}\left[\partial_{x}\widetilde{B}^{y}-\partial_{y}\widetilde{B}^{x}\right]|^{n}-d\tau\frac{\tau_{n}}{\tau_{n+1/2}}\,J^{\eta}_{m}|^{n}.
	\end{split}
\end{align}
Here the electric fields are defined at time $\tau = \tau_0 + (n\pm 1/2) d\tau$, while the magnetic fields are defined at time $\tau = \tau_0 + n d\tau$. Regarding Ohm's law in the medium current $J^{\mu}_{m}$, the aforementioned set of equations can be further expressed as follows,
\begin{align}\label{app-E}
	\begin{split}
		&\widetilde{E}^{x}|^{n+1/2}=COE|^{n}\frac{\tau_{n-1/2}}{\tau_{n+1/2}}\widetilde{E}^{x}|^{n-1/2}+d\tau\frac{CE|^{n}}{\tau_{n+1/2}}\left[\partial_{y}(\tau^{2}\widetilde{B}^{\eta})-\partial_{\eta}\widetilde{B}^{y}\right]|^{n}-d\tau\,CE|^{n}\frac{\tau_{n}}{\tau_{n+1/2}}\,\widetilde{J}^{x}_{m}|^{n},\\
		&\widetilde{E}^{y}|^{n+1/2}=COE|^{n}\frac{\tau_{n-1/2}}{\tau_{n+1/2}}\widetilde{E}^{y}|^{n-1/2}-d\tau\,\frac{CE|^{n}}{\tau_{n+1/2}}\left[\partial_{x}(\tau^{2}\widetilde{B}^{\eta})-\partial_{\eta}\widetilde{B}^{x}\right]|^{n}-d\tau\,CE|^{n}\frac{\tau_{n}}{\tau_{n+1/2}}\,\widetilde{J}^{y}_{m}|^{n},\\
		&\widetilde{E}^{\eta}|^{n+1/2}=COE|^{n}\frac{\tau_{n-1/2}}{\tau_{n+1/2}}\widetilde{E}^{\eta}|^{n-1/2}+d\tau\,\frac{CE|^{n}}{\tau_{n+1/2}}\left[\partial_{x}\widetilde{B}^{y}-\partial_{y}\widetilde{B}^{x}\right]|^{n}-d\tau\,CE|^{n}\frac{\tau_{n}}{\tau_{n+1/2}}\,\widetilde{J}^{\eta}_{m}|^{n}.
	\end{split}
\end{align}
In this derivation, we utilized the approximation $\tau\,E^{i}|^{n}=(\tau_{n+1/2}E^{i}|^{n+1/2}+\tau_{n-1/2}E^{i}|^{n-1/2})/2$. The newly introduced coefficients are defined as 
\begin{align}
	&CE|^{n}=\frac{2}{2+d\tau\,\sigma\,u^{\tau}|^{n}},\qquad COE|^{n}=\frac{2-d\tau\,\sigma\,u^{\tau}|^{n}}{2+d\tau\,\sigma\,u^{\tau}|^{n}}.
\end{align}
And the new current is expressed as follows,
\begin{align}
\begin{split}
&\widetilde{J}^{x}_{m}=n_{q}\,u^{x}+d_{q}^{x}+\sigma\,(\tau\widetilde{B}^{\eta}u^{y}-\widetilde{B}^{y}\tau\,u^{\eta})+\sigma\,F^{x\nu}_{M,ext}u_{\nu},\\
&\widetilde{J}^{y}_{m}=n_{q}\,u^{y}+d_{q}^{y}+\sigma\,(-\tau\widetilde{B}^{\eta}u^{x}+\widetilde{B}^{x}\tau\,u^{\eta})+\sigma\,F^{y\nu}_{M,ext}u_{\nu},\\
&\widetilde{J}^{\eta}_{m}=n_{q}\,u^{\eta}+d_{q}^{\eta}+\sigma\,(\widetilde{B}^{y}u^{x}/\tau-\widetilde{B}^{x}u^{y}/\tau)+\sigma\,F^{\eta\nu}_{M,ext}u_{\nu},\\
\end{split}
\end{align}

To Eq. (\ref{eq:ME-Milne-B}), the algorithm can be written as follows,
\begin{align}
	\begin{split}
		&\partial_{\tau}(\tau\,\widetilde{B}^{x})|^{n+1/2}=-\partial_{y}(\tau^{2}\widetilde{E}^{\eta})|^{n+1/2}+\partial_{\eta}\widetilde{E}^{y}|^{n+1/2},\\
		&\partial_{\tau}(\tau\,\widetilde{B}^{y})|^{n+1/2}=\partial_{x}(\tau^{2}\widetilde{E}^{\eta})|^{n+1/2}-\partial_{\eta}\widetilde{E}^{x}|^{n+1/2},\\
		&\partial_{\tau}(\tau\,\widetilde{B}^{\eta})|^{n+1/2}=-\partial_{x}\widetilde{E}^{y}|^{n+1/2}+\partial_{y}\widetilde{E}^{x}|^{n+1/2}.
	\end{split}
\end{align}
Similar to the aforementioned approach, employing the finite center difference method for the time derivative results in
\begin{align}\label{app-B}
	\begin{split}
		&\widetilde{B}^{x}|^{n+1}=\frac{\tau_{n}}{\tau_{n+1}}\widetilde{B}^{x}|^{n}-\frac{d\tau}{\tau_{n+1}}\Big[\partial_{y}(\tau^{2}\widetilde{E}^{\eta})-\partial_{\eta}\widetilde{E}^{y}\Big]|^{n+1/2},\\
		&\widetilde{B}^{y}|^{n+1}=\frac{\tau_{n}}{\tau_{n+1}}\widetilde{B}^{y}|^{n}+\frac{d\tau}{\tau_{n+1}}\Big[\partial_{x}(\tau^{2}\widetilde{E}^{\eta})-\partial_{\eta}\widetilde{E}^{x}\Big]|^{n+1/2},\\
		&\widetilde{B}^{\eta}|^{n+1}=\frac{\tau_{n}}{\tau_{n+1}}\widetilde{B}^{\eta}|^{n}-\frac{d\tau}{\tau_{n+1}}\Big[\partial_{x}\widetilde{E}^{y}-\partial_{y}\widetilde{E}^{x}\Big]|^{n+1/2}.
	\end{split}
\end{align}
For now, we are solely focusing on the temporal aspect of the scheme. Next, let's delve into the spatial component. As introduced by Yee, the electric and magnetic fields are defined on staggered spatial grids, i.e 
\begin{align}
\begin{split}
&\widetilde{E}^{x}|^{n+1/2}_{i+1/2,j,k}, \widetilde{E}^{y}|^{n+1/2}_{i,j+1/2,k}, \widetilde{E}^{z}|^{n+1/2}_{i,j,k+1/2},\\
&\widetilde{B}^{x}|^{n}_{i,j+1/2,k+1/2}, \widetilde{B}^{y}|^{n}_{i+1/2,j,k+1/2}, \widetilde{B}^{\eta}|^{n}_{i+1/2,j+1/2,k}.\\
\end{split}
\end{align}
The spatial derivative will also be computed using the finite center difference method. The set of equations in Eq. (\ref{app-E}) is then transformed into the following form
\begin{align}
	\begin{split}
		&\widetilde{E}^{x}|^{n+1/2}_{i_h,j,k}=COE|^{n}_{i_h,j,k}\frac{\tau_{n-1/2}}{\tau_{n+1/2}}\widetilde{E}^{x}|^{n-1/2}_{i_h,j,k}-d\tau\,CE|^{n}_{i_h,j,k}\frac{\tau_{n}}{\tau_{n+1/2}}\,\widetilde{J}^{x}_{m}|^{n}_{i_h,j,k}\\
		&~~~~~~~~~~~~+d\tau\,\frac{CE|^{n}_{i_h,j,k}}{\tau_{n+1/2}}\left[\tau^{2}_n\frac{\widetilde{B}^{\eta}|^{n}_{i_h,j_h,k}-\widetilde{B}^{\eta}|^{n}_{i_h,j_b,k}}{dy}-\frac{\widetilde{B}^{y}|^{n}_{i_h,j,k_h}-\widetilde{B}^{y}|^{n}_{i_h,j,k_b}}{d\eta}\right],\\
		&\widetilde{E}^{y}|^{n+1/2}_{i,j_h,k}=COE|^{n}_{i,j_h,k}\frac{\tau_{n-1/2}}{\tau_{n+1/2}}\widetilde{E}^{y}|^{n-1/2}_{i,j_h,k}-d\tau\,CE|^{n}_{i,j_h,k}\frac{\tau_{n}}{\tau_{n+1/2}}\,\widetilde{J}^{y}_{m}|^{n}_{i,j_h,k}\\
		&~~~~~~~~~~~~-d\tau\,\frac{CE|^{n}_{i,j_h,k}}{\tau_{n+1/2}}\left[\tau^{2}_n\frac{\widetilde{B}^{\eta}|^{n}_{i_h,j_h,k}-\widetilde{B}^{\eta}|^{n}_{i_b,j_h,k}}{dx}-\frac{\widetilde{B}^{x}|^{n}_{i,j_h,k_h}-\widetilde{B}^{x}|^{n}_{i,j_h,k_b}}{d\eta}\right],\\
		&\widetilde{E}^{\eta}|^{n+1/2}_{i,j,k_h}=COE|^{n}_{i,j,k_h}\frac{\tau_{n-1/2}}{\tau_{n+1/2}}\widetilde{E}^{\eta}|^{n-1/2}_{i,j,k_h}-d\tau\,CE|^{n}_{i,j,k_h}\frac{\tau_{n}}{\tau_{n+1/2}}\,\widetilde{J}^{\eta}_{m}|^{n}_{i,j,k_h}\\
		&~~~~~~~~~~~~+d\tau\,\frac{CE|^{n}_{i,j,k_h}}{\tau_{n+1/2}}\left[\frac{\widetilde{B}^{y}|^{n}_{i_h,j,k_h}-\widetilde{B}^{y}|^{n}_{i_b,j,k_h}}{dx}-\frac{\widetilde{B}^{x}|^{n}_{i,j_h,k_h}-\widetilde{B}^{x}|^{n}_{i,j_b,k_h}}{dy}\right].
	\end{split}
\end{align}
The symbols $i_h = i + 1/2$ and $i_b = i - 1/2$ are defined, and the same convention is applied for $j$ and $k$. It should be noted that the magnetic fields in the aforementioned corresponding currents are not located at the same position as the computed magnetic field. Consequently, it becomes necessary to represent them in conjunction with the computed magnetic field, which can be approximated as follows,
\begin{align}
	\begin{split}
		&\widetilde{B}^{y}|^{n}_{i_h,j,k}=\Big(\widetilde{B}^{y}|^{n}_{i_h,j,k_h}+\widetilde{B}^{y}|^{n}_{i_h,j,k_b}\Big)/2,~~\widetilde{B}^{\eta}|^{n}_{i_h,j,k}=\Big(\widetilde{B}^{\eta}|^{n}_{i_h,j_h,k}+\widetilde{B}^{\eta}|^{n}_{i_h,j_b,k}\Big)/2,\\
		&\widetilde{B}^{x}|^{n}_{i,j_h,k}=\Big(\widetilde{B}^{x}|^{n}_{i,j_h,k_h}+\widetilde{B}^{x}|^{n}_{i,j_h,k_b}\Big)/2,~~\widetilde{B}^{\eta}|^{n}_{i,j_h,k}=\Big(\widetilde{B}^{\eta}|^{n}_{i_h,j_h,k}+\widetilde{B}^{\eta}|^{n}_{i_b,j_h,k}\Big)/2,\\
		&\widetilde{B}^{x}|^{n}_{i,j,k_h}=\Big(\widetilde{B}^{x}|^{n}_{i,j_h,k_h}+\widetilde{B}^{x}|^{n}_{i,j_b,k_h}\Big)/2,~~\widetilde{B}^{y}|^{n}_{i,j,k_h}=\Big(\widetilde{B}^{y}|^{n}_{i_h,j,k_h}+\widetilde{B}^{y}|^{n}_{i_b,j,k_h}\Big)/2.\\
	\end{split}
\end{align}

Similarly, Eq.(\ref{app-B}) now can be written as,
\begin{align}
	\begin{split}
		&\widetilde{B}^{x}|^{n+1}_{i,j_h,k_h}=\frac{\tau_{n}}{\tau_{n+1}}\widetilde{B}^{x}|^{n}_{i,j_h,k_h}-\frac{d\tau}{\tau_{n+1}}\Big[\tau^{2}_{n+1/2}\frac{\widetilde{E}^{\eta}|^{n+1/2}_{i,j+1,k_h}-\widetilde{E}^{\eta}|^{n+1/2}_{i,j,k_h}}{dy}-\frac{\widetilde{E}^{y}|^{n+1/2}_{i,j_h,k+1}-\widetilde{E}^{y}|^{n+1/2}_{i,j_h,k}}{d\eta}\Big],\\
		&\widetilde{B}^{y}|^{n+1}_{i_h,j,k_h}=\frac{\tau_{n}}{\tau_{n+1}}\widetilde{B}^{y}|^{n}_{i_h,j,k_h}+\frac{d\tau}{\tau_{n+1}}\Big[\tau^{2}_{n+1/2}\frac{\widetilde{E}^{\eta}|^{n+1/2}_{i+1,j,k_h}-\widetilde{E}^{\eta}|^{n+1/2}_{i,j,k_h}}{dx}-\frac{\widetilde{E}^{x}|^{n+1/2}_{i_h,j,k+1}-\widetilde{E}^{x}|^{n+1/2}_{i_h,j,k}}{d\eta}\Big],\\
		&\widetilde{B}^{\eta}|^{n+1}_{i_h,j_h,k}=\frac{\tau_{n}}{\tau_{n+1}}\widetilde{B}^{\eta}|^{n}_{i_h,j_h,k}-\frac{d\tau}{\tau_{n+1}}\Big[\frac{\widetilde{E}^{y}|^{n+1/2}_{i+1,j_h,k}-\widetilde{E}^{y}|^{n+1/2}_{i,j_h,k}}{dx}-\frac{\widetilde{E}^{x}|^{n+1/2}_{i_h,j+1,k}-\widetilde{E}^{x}|^{n+1/2}_{i_h,j,k}}{dy}\Big].
	\end{split}
\end{align}

\section{Levi--Civita tensor and electromagnetic tensor in Milne space}\label{ap:levi-civita}
In Milne coordinate, the Levi--Civita tensor is written as the following\cite{Huang:2022qdn},
\begin{align}\begin{split}
		&\epsilon^{\mu\nu\rho\sigma}_{M}=\frac{1}{\sqrt{|g|}}\epsilon^{\mu\nu\rho\sigma}
		,\qquad
		\epsilon_{\mu\nu\rho\sigma}^{M}=\theta(g)\sqrt{|g|}\epsilon_{\mu\nu\rho\sigma}
		,
\end{split}\end{align}
In this expression, $g=\mathrm{det}(g_{\mu\nu})$, $\theta$ represents the Heaviside step function, and the Levi-Civita symbol $\widetilde{\epsilon}^{\mu\nu\rho\sigma}$ is defined in Minkowski space as follows,
\begin{align*}
	&\epsilon^{\mu\nu\rho\sigma}=\epsilon_{\mu\nu\rho\sigma}=\left\{
	\begin{array}{rl}
	    +1,&\qquad \text{even permutation of (0,1,2,...n-1)}, \\
		-1,&\qquad \text{odd permutation of (0,1,2,...n-1)},\\
	    0,&\qquad \text{otherwise}.
	\end{array}
	\right.
\end{align*}
By setting $g=-\tau^{2}$ and $|g|=\tau^{2}$, we obtain $\epsilon^{0123}=\frac{1}{\tau}$ and $\epsilon_{0123}=-\tau$.

In accordance with the following relations between electric and magnetic fields and the electromagnetic tensor:
\begin{align}
	&\widetilde{E}^{i}=F_{M}^{i0},\qquad \widetilde{B}^{i}=\widetilde{F}_{M}^{i0}.
\end{align}
one can directly derive the electromagnetic tensor in Milne space using the aforementioned Levi-Civita definition. This can be expressed as follows,
\begin{align}
	&F^{\mu\nu}_{M}=\begin{pmatrix}
		0 & -\widetilde{E}^{x} & -\widetilde{E}^{y} & -\widetilde{E}^{\eta}\\
		\widetilde{E}^{x} & 0 & -\tau\,\widetilde{B}^{\eta} &\frac{\widetilde{B}^{y}}{\tau}\\
		\widetilde{E}^{y} &\tau\,\widetilde{B}^{\eta} & 0 & -\frac{\widetilde{B}^{x}}{\tau} \\
		\widetilde{E}^{\eta} & -\frac{\widetilde{B}^{y}}{\tau} &\frac{\widetilde{B}^{x}}{\tau} &0 \\
	\end{pmatrix}, \qquad 	&\widetilde{F}^{\mu\nu}_{M}=\begin{pmatrix}
		0 & -\widetilde{B}^{x} & -\widetilde{B}^{y} & -\widetilde{B}^{\eta}\\
		\widetilde{B}^{x} & 0 & \tau\,\widetilde{E}^{\eta} &-\frac{\widetilde{E}^{y}}{\tau}\\
		\widetilde{B}^{y} &-\tau\,\widetilde{E}^{\eta} & 0 & \frac{\widetilde{E}^{x}}{\tau} \\
		\widetilde{B}^{\eta} & \frac{\widetilde{E}^{y}}{\tau} &-\frac{\widetilde{E}^{x}}{\tau} &0 \\
	\end{pmatrix}
\end{align}

\end{widetext}

\end{appendix}

\bibliography{reference}
\bibliographystyle{unsrt}

\end{document}